\documentclass[letterpaper,twocolumn,10pt]{article}
\usepackage{usenix}

\usepackage[table]{xcolor}  
\usepackage{array}
\usepackage{xspace}
\usepackage{graphicx}
\usepackage{tabularx}
\usepackage{booktabs}
\usepackage{tikz}  
\usetikzlibrary{positioning,shapes.geometric,arrows.meta,calc,backgrounds,decorations.markings}
\usepackage{amssymb}
\usepackage{pifont}
\usepackage{url}
\usepackage{hyperref}
\usepackage{microtype}
\microtypecontext{spacing=nonfrench}
\usepackage{mdframed}

\newcommand{\stix}{STIX\xspace}
\newcommand{\mitre}{MITRE\xspace}
\newcommand{\attack}{ATT\&CK\xspace}
\newcommand{\caldera}{Caldera\xspace}

\newcommand{\nIntrusionSets}{172}
\newcommand{\nRelationships}{17270}
\newcommand{\nAttackPatternObjects}{691}
\newcommand{\nTechniquesAnalysis}{691}

\newcommand{\campaignCoveragePct}{43.0}

\newcommand{\campaignTechniqueCount}{297}
\newcommand{\intrusionSetCoveragePct}{70.6}
\newcommand{\intrusionSetTechniqueCount}{488}
\newcommand{\sharedTechniqueCount}{268}
\newcommand{\campaignOnlyCount}{29}
\newcommand{\intrusionSetOnlyCount}{220}
\newcommand{\unusedTechniqueCount}{174}

\newcommand{\campaignOnlyPct}{4.2}
\newcommand{\sharedPct}{38.8}
\newcommand{\intrusionSetOnlyPct}{31.8}
\newcommand{\unusedPct}{25.2}

\newcommand{\nIntrusionSetsWithTechniques}{168}
\newcommand{\intrusionSetWithTechniquesPct}{97.7}
\newcommand{\nIntrusionSetsEmpty}{4}
\newcommand{\intrusionSetEmptyPct}{2.3}
\newcommand{\intrusionSetTechRefs}{4362}
\newcommand{\intrusionSetTechAvg}{26.0}

\newcommand{\nPlatformAgnosticTechniques}{32}
\newcommand{\nTranslatableTechniques}{659}

\newcommand{\nAppendixAttackPatterns}{691}
\newcommand{\AppendixFieldPopulationRows}{%
\texttt{kill\_chain\_phases} & 691 & 691 \\
\texttt{x\_mitre\_platforms} & 691 & 691 \\
\texttt{x\_mitre\_system\_requirements} & 0 & 0 \\
\texttt{x\_mitre\_detection} & 691 & 0 \\
\texttt{x\_mitre\_data\_sources} & 0 & 0 \\
\texttt{x\_mitre\_permissions\_required} & 0 & 0 \\
}
\newcommand{\AppendixItemsetRows}{%
1 & 28 & 54.9\% \\
2 & 17 & 33.3\% \\
3 & 10 & 19.6\% \\
4 & 7 & 13.7\% \\
5 & 6 & 11.8\% \\
}

\definecolor{execblue}{HTML}{4477AA}   
\definecolor{ctired}{HTML}{EE6677}     
\definecolor{loopgreen}{HTML}{228833}  
\definecolor{tolyellow}{HTML}{CCBB44}  
\definecolor{tolcyan}{HTML}{66CCEE}    
\definecolor{tolpurple}{HTML}{AA3377}  
\definecolor{edgegray}{HTML}{333333}   
\definecolor{midgray}{HTML}{666666}    
\definecolor{dashgray}{HTML}{BBBBBB}   
\definecolor{stagefill}{HTML}{F5F5F5}  
\definecolor{facegray}{HTML}{FFFFFF}   
\definecolor{findBlue}{HTML}{4477AA}
\definecolor{findBg}{HTML}{EEF3FA}

\tikzset{
  computer/.pic={
    \draw[fill=dashgray!20, draw=edgegray!60, rounded corners=1.5pt,
          line width=0.5pt]
      (-0.55,-0.45) rectangle (0.55,0.45);
    \fill[dashgray!30] (-0.46,-0.36) rectangle (0.46,0.36);
    \draw[fill=edgegray!25, draw=edgegray!60, line width=0.5pt]
      (-0.12,-0.63) rectangle (0.12,-0.45);
    \draw[fill=edgegray!20, draw=edgegray!60, line width=0.5pt]
      (-0.33,-0.70) rectangle (0.33,-0.63);
  }
}

\newmdenv[
  topline=false, bottomline=false, rightline=false,
  linewidth=2.5pt, linecolor=findBlue,
  backgroundcolor=findBg,
  innerleftmargin=7pt, innerrightmargin=5pt,
  innertopmargin=4pt, innerbottommargin=4pt,
  skipabove=5pt, skipbelow=3pt
]{findingenv}

\newmdenv[
  topline=true, bottomline=true, leftline=false, rightline=false,
  linewidth=0.6pt, linecolor=black!30,
  backgroundcolor=gray!5,
  innerleftmargin=6pt, innerrightmargin=6pt,
  innertopmargin=5pt, innerbottommargin=5pt,
  skipabove=6pt, skipbelow=4pt
]{textbox}

\AtBeginDocument{%
  }

\author{
{\rm Ágney Lopes Roth Ferraz}\thanks{These authors contributed equally.}\\
Aeronautics Institute of Technology\\
{\tt roth@ita.br}
\and
{\rm Sidnei Barbieri}\footnotemark[1]\\
Aeronautics Institute of Technology \&\\
Carnegie Mellon University\\
{\tt sbarbier@andrew.cmu.edu}
\and
{\rm Wagner Comin Sonaglio}\\
Aeronautics Institute of Technology\\
{\tt sonaglio@ita.br}
\and
{\rm Lourenço Alves Pereira Júnior}\\
Aeronautics Institute of Technology\\
{\tt ljr@ita.br}
}

\title{The Procedural Semantics Gap in ATT\&CK-in-STIX: A Measurement-Driven Analysis for APT Emulation}
\begin{document}
\maketitle

\begin{abstract}
Public cyber threat intelligence (CTI) represented in Structured Threat Information Expression (\stix{}) has become the de facto standard for sharing adversary knowledge, with \mitre\attack{} serving as its most widely adopted knowledge base. While this representation effectively captures adversary tactics and techniques, it was designed to describe observed behavior rather than to specify executable attack procedures. This raises a fundamental question: does public \attack{}-in-\stix{} provide sufficient behavioral information to support emulation of multi-stage adversaries? To answer this question, we perform a measurement study of the \attack{} Enterprise dataset and analyze the behavioral information encoded in campaign and intrusion-set objects. Our analysis shows that campaign descriptions contain only fragmented behavioral evidence: only \campaignCoveragePct\% of ATT\&CK techniques appear in at least one campaign, and neither clustering nor Longest Common Subsequence (LCS)-based analyses reveal reusable procedural structures. Although intrusion sets provide broader coverage of techniques, they likewise omit the execution order, operational dependencies, and environmental assumptions required to execute attacks. These findings demonstrate that public CTI largely specifies \emph{what} adversaries do while leaving much of \emph{how} those behaviors should be executed undefined.

Motivated by this gap, we propose a three-stage methodology for translating structured CTI into executable adversary operations and implement it within the \mitre{} \caldera{} framework. Case studies based on the \emph{ShadowRay} and \emph{Soft Cell} campaigns demonstrate that reliable multi-stage emulation is only possible after the missing procedural information is explicitly reconstructed and validated. We further audit the eight curated adversaries distributed with the legacy Docker artifact and verify that all workflows terminate successfully, leaving no residual active links. Together, our results establish a reproducible boundary between the behavioral knowledge encoded in public CTI and the procedural semantics required to achieve automated adversary emulation.
\end{abstract}

\section{Introduction}
\label{sec:introduction}

Advanced Persistent Threats (APTs) conduct coordinated, multi-stage operations that evolve over long periods, adapt to changing environments, and combine multiple techniques across hosts to achieve operational objectives~\cite{barr-smith2021survivalism,lourenco2024devil,milajerdi2019holmes}. As organizations increasingly adopt continuous security validation, adversary emulation has become an important mechanism for evaluating defensive capabilities against realistic attack behavior rather than isolated techniques.

Public Cyber Threat Intelligence (CTI) appears to be a natural foundation for this purpose. Modern CTI standards describe campaigns, intrusion sets, malware, tools, and ATT\&CK techniques in machine-readable form, enabling organizations to exchange structured knowledge about real adversaries~\cite{apurva2024mitre,strom2018mitre,stixspec}. Among these standards, the \mitre{} \attack{} Enterprise knowledge base, distributed as Structured Threat Information Expression (\stix{}) bundles, has become the de facto public representation of adversary behavior.

However, ATT\&CK-in-STIX was designed to document observed behavior rather than executable attack procedures. Campaign and intrusion-set objects identify techniques and behavioral relationships but generally omit execution order, operational dependencies, parameter values, privilege transitions, and environmental assumptions required to reproduce multi-stage attacks~\cite{stixspec,10117505,savat2021extractor}. Consequently, although structured CTI is widely used for threat intelligence, ATT\&CK mapping, and defensive coverage analysis, its suitability for direct emulation of adversaries remains largely unexplored.

Existing research addresses adjacent problems but not this question. Provenance-based approaches reconstruct attack progression from execution telemetry~\cite{hossain202comb,10646673,10646725,zian2024magic}, while CTI extraction systems transform narrative reports into structured threat representations~\cite{savat2021extractor,xu2024intelex,cheng2025ctinexus}. Although these directions improve attack reconstruction and CTI usability, neither evaluates whether the procedural information already contained in public ATT\&CK-in-STIX is sufficient to support reproducible multi-stage adversary emulation without introducing external knowledge.

\begin{figure}[t]
    \centering
    \includegraphics[width=0.99\linewidth]{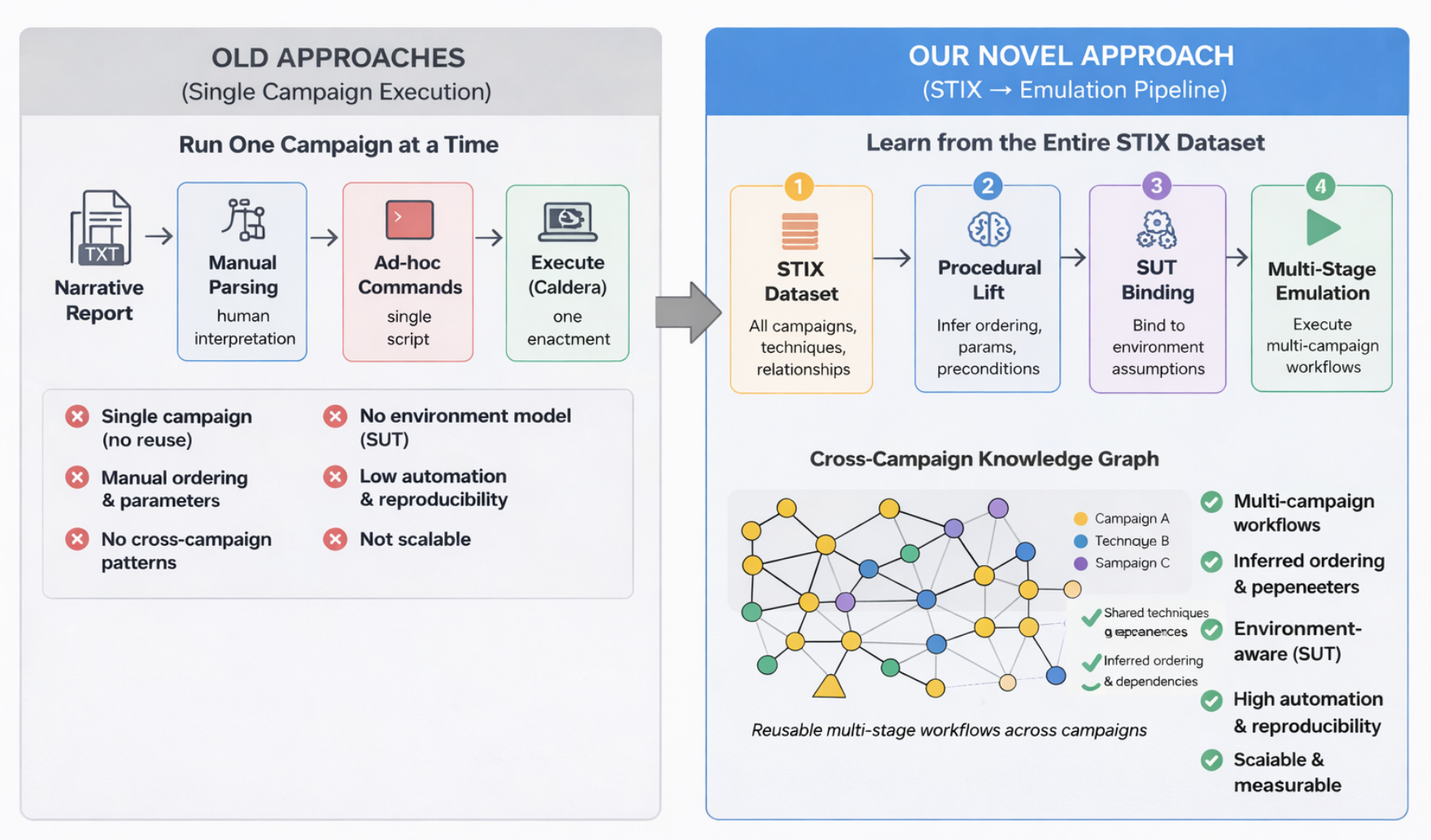}
    \caption{From descriptive CTI to executable adversary emulation. Traditional approaches reproduce individual campaigns, whereas our methodology translates structured STIX datasets into reusable multi-stage adversary workflows through procedural inference and environment-aware binding.}
    \label{fig:novelty}
\end{figure}

In this paper, we investigate the procedural sufficiency of public ATT\&CK-in-STIX for adversary emulation. We focus on the ATT\&CK Enterprise bundle because it provides the richest publicly available collection of campaigns, intrusion sets, techniques, and semantic relationships within a single coherent corpus. We first perform a measurement study to quantify the behavioral information encoded in campaign and intrusion-set objects through coverage, overlap, clustering, and sequence analyses. Based on the observed limitations, we then propose a three-stage translation methodology that explicitly separates information recovered from structured CTI from the procedural knowledge required for execution. The methodology is implemented using the \mitre{} \caldera{} adversary emulation framework and evaluated through representative campaign case studies. Figure~\ref{fig:novelty} summarizes this transition from descriptive CTI to executable adversary emulation.

Our measurements show that public ATT\&CK-in-STIX captures substantial behavioral knowledge but does not encode reusable executable procedures. Campaigns cover only a subset of ATT\&CK techniques, behavioral similarity across campaigns is weak, and neither clustering nor sequence analyses reveals common procedural structures suitable for direct execution. These findings establish a clear gap in procedural semantics between descriptive CTI and executable adversary workflows.

This study is guided by the following research questions.

\noindent\hangindent=2.7em\hangafter=1
\textbf{RQ1:} How much adversarial behavior do current ATT\&CK-in-STIX artifacts capture for multi-stage APT emulation?

\noindent\hangindent=2.7em\hangafter=1
\textbf{RQ2:} How can structured CTI be transformed into computable multi-stage behavioral plans for adversary emulation?

\noindent\hangindent=2.7em\hangafter=1
\textbf{RQ3:} Which structural and semantic gaps in descriptive CTI prevent automation, and how can these limitations be systematically mitigated?

Our contributions are summarized as follows.

\begin{itemize}

\item We present the first systematic measurement of procedural sufficiency in ATT\&CK-in-STIX by quantifying coverage, sparsity, overlap, clustering behavior, and sequence similarity within the ATT\&CK Enterprise bundle.

\item We identify the gap in procedural semantics between structured CTI and executable adversary emulation, showing that execution order, operational dependencies, parameterization, and environment-specific assumptions are largely absent from public STIX representations.

\item We propose a three-stage translation methodology that explicitly distinguishes automatically recoverable information from analyst-supplied procedural knowledge, thereby establishing reproducible boundaries for CTI-driven adversary emulation.

\item We validate the proposed methodology through an end-to-end implementation in the \mitre{} \caldera{} framework using representative campaign case studies and curated adversary workflows, demonstrating that successful execution depends on explicitly reconstructing the missing procedural semantics rather than on the descriptive completeness of structured CTI.

\end{itemize}

\section{Background}
\label{sec:background}

Cyber Threat Intelligence (CTI) provides structured representations of adversaries, their capabilities, and observed operations. To improve interoperability, the cybersecurity community adopted Structured Threat Information Expression (\stix{}), a machine-readable standard for exchanging threat intelligence across organizations~\cite{stixspec}. Rather than representing executable attack procedures, STIX models threat knowledge as interconnected entities, such as campaigns, intrusion sets, malware, tools, and attack patterns, along with the semantic relationships among them.

The \mitre{} \attack{} framework provides a curated knowledge base of adversary tactics and techniques observed in real-world operations~\cite{strom2018mitre}. ATT\&CK defines the behavioral vocabulary, while STIX provides the representation for distributing that knowledge. The ATT\&CK Enterprise bundle combines techniques, campaigns, intrusion sets, malware, tools, mitigations, and semantic relationships into a single STIX dataset, making it one of the most widely adopted public CTI resources for threat-informed defense and adversary emulation~\cite{jin2024sharing,strom2018mitre}.

For adversary emulation, the most relevant ATT\&CK entities are campaigns and intrusion sets. Campaigns describe individual operations by associating observed techniques with a specific intrusion, whereas intrusion sets summarize recurring behaviors observed across multiple campaigns~\cite{abdulellah2021atlas}. Techniques are represented as \texttt{attack-pattern} objects, while relationship objects capture associations such as \texttt{uses}, \texttt{attributed-to}, and \texttt{subtechnique-of}. Together, these objects form a graph describing observed adversary behavior.

\begin{figure}[t]
    \centering
    \includegraphics[width=0.99\linewidth]{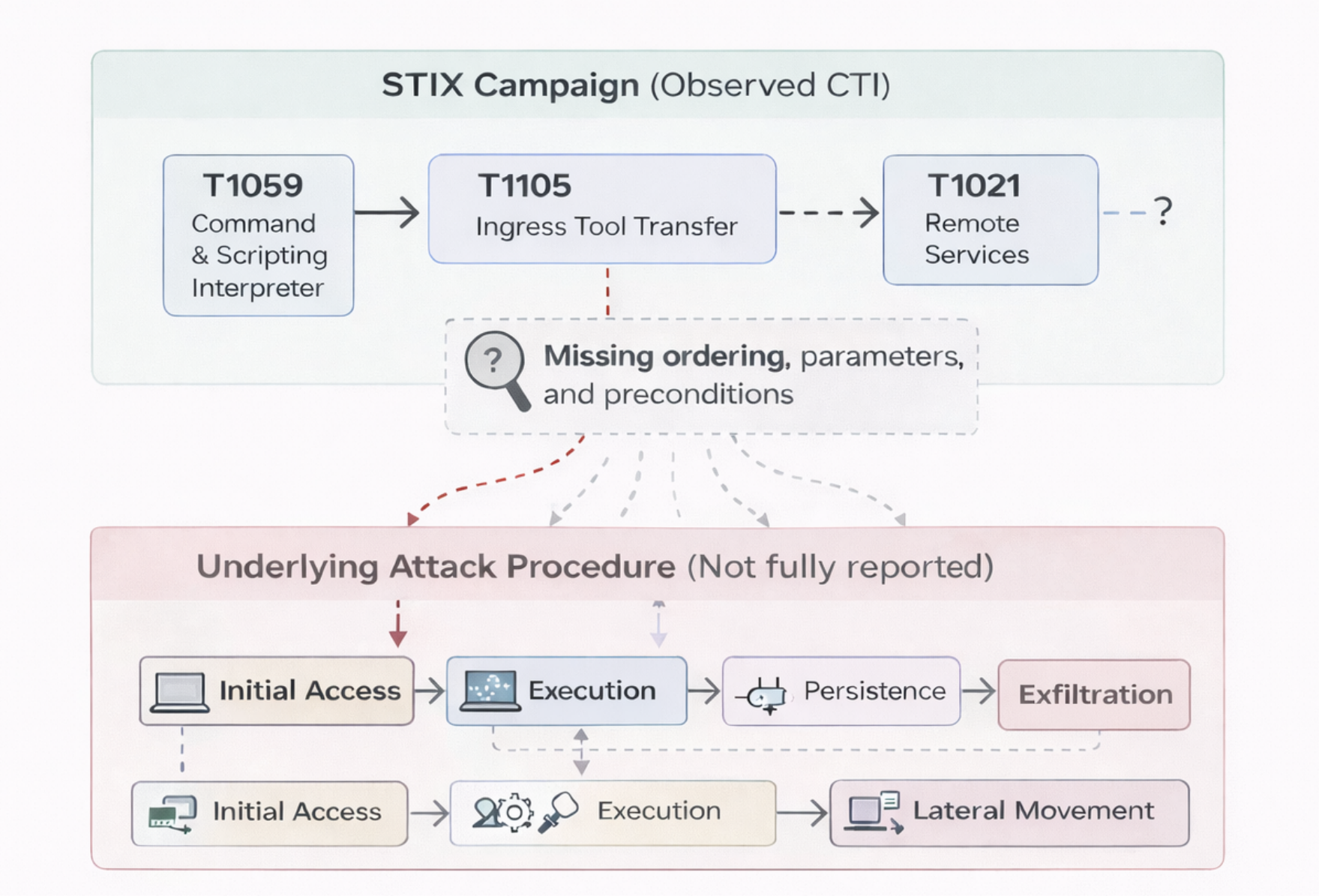}
    \caption{Structured CTI describes adversary behavior through techniques and semantic relationships but omits the procedural information required for execution, including ordering, dependencies, parameterization, and environmental assumptions.}
    \label{fig:gap}
\end{figure}

Although ATT\&CK-in-STIX captures rich behavioral knowledge, it does not explicitly encode executable procedures. Relationships indicate which techniques, malware, or tools are associated with an adversary, but they do not specify execution order, operational dependencies, parameter values, privilege transitions, or environmental assumptions~\cite{stixspec}. Consequently, multiple executable workflows may be consistent with the same structured CTI representation.

This distinction is particularly important for adversary emulation. Frameworks such as \caldera{} execute ordered attack workflows composed of concrete commands, execution constraints, environmental bindings, and cleanup procedures~\cite{apurva2024mitre,10.1145/3634737.3645012,singer2025incalmo}. These procedural elements are not represented directly in ATT\&CK-in-STIX and therefore must be supplied during translation. The difference between the behavioral knowledge encoded in structured CTI and the procedural information required for execution is referred to in this paper as the \emph{gap in procedural semantics}.

\begin{table*}[!t]
  \caption{\stix{} sources in the consolidated dataset.}
  \label{tab:sources}
  \centering
  \small
\resizebox{\textwidth}{!}{%
  \begin{tabular}{lccccl}
    \toprule
    Source & Type & Objects & Unique & Duplicates & Description \\
    \midrule
    \rowcolor{gray!10}
    \mitre{} Enterprise dataset & Repository & 24,771 & 24,771 & 0.0\% & Broad collection including TTPs, adversaries, and campaigns. \\
    \mitre{} CAPEC & Repository & 2,666 & 2,666 & 0.0\% & Attack patterns. \\
    \rowcolor{gray!10}
    DigitalSide & Repository & 38,306 & 38,306 & 0.0\% & Indicators, techniques, relationships, and malware with IOC focus. \\
    AlienVault OTX & TAXII server & 17,303 & 17,303 & 0.0\% & Diverse objects including \texttt{attack-pattern} and \texttt{threat-actor}. \\
    \rowcolor{gray!10}
    EclecticIQ & TAXII server & 2,335 & 2,043 & 12.5\% & Threat actors, campaigns, and reports. \\
    \midrule
    \textbf{Total (deduplicated)} & & \textbf{85,381} & \textbf{85,089} & 0.0\% & Nearly duplicate-free consolidated dataset. \\
    \bottomrule
  \end{tabular}
}
\end{table*}

\section{Related Work}
\label{sec:related}

Our work is related to four research directions: structured cyber threat intelligence, CTI extraction, attack reconstruction, and adversary emulation. While these areas have advanced the representation and operational use of cyber threat intelligence, none directly addresses whether public ATT\&CK-in-STIX contains sufficient procedural information to support executable multi-stage adversary emulation.

A substantial body of work has investigated the quality, consistency, and interoperability of structured CTI representations. Previous studies analyze STIX data models, ATT\&CK annotations, commercial threat-sharing platforms, and detection-rule ecosystems, reporting heterogeneous field usage, uneven coverage, inconsistent labeling, and vendor-specific reporting practices~\cite{stixspec,jin2024sharing,10117505,strom2018mitre,apurva2024mitre,10.1145/3634737.3645012,vermeer2022evolution}. These studies establish that public CTI exhibits significant variability across providers and that structured intelligence should not be assumed to be complete or uniform.

Our objective differs fundamentally from these analyses. Rather than evaluating data quality or interoperability, we investigate whether the behavioral information already encoded in public ATT\&CK-in-STIX is sufficient for procedural emulation of adversaries. In particular, we study whether campaign and intrusion-set objects contain enough information to reconstruct executable attack workflows without introducing additional procedural knowledge.

A recent study by Saha \emph{et al.}~\cite{saha2026kitten} is particularly relevant because it evaluates the distinctiveness of ATT\&CK threat-group profiles for attribution. Their work demonstrates that techniques alone rarely distinguish adversaries and that software usage often provides stronger attribution signals. Our work addresses a complementary question. Instead of measuring attribution specificity, we measure procedural sufficiency by asking whether the behavioral structure encoded in ATT\&CK is sufficiently detailed to reproduce multi-stage adversary behavior.

Another line of research focuses on transforming unstructured threat reports into structured CTI. Systems such as Extractor~\cite{savat2021extractor}, CTI HAL~\cite{dellapenna2025ctihal}, and subsequent NLP- and LLM-based approaches recover ATT\&CK techniques, malware, infrastructure, and semantic relationships from narrative reports~\cite{gabrys2024nlp,marvin2025sokttp,xu2024intelex,cheng2025ctinexus,deng2024raconteur}. These methods substantially improve the scalability of CTI construction and enable the creation of richer knowledge graphs from previously unstructured intelligence.

However, these approaches evaluate extraction accuracy rather than executable completeness. Temporal ordering, operational dependencies, execution constraints, and environmental assumptions are generally inferred from narrative text or introduced during post-processing rather than being measured directly from the STIX representation itself. Consequently, they do not establish whether public structured CTI already includes sufficient procedural semantics to support emulation of adversaries.

A separate body of work reconstructs attack progression from system telemetry. Provenance-based systems such as HOLMES, UNICORN, ATLAS, ALchemist, and NODLINK~\cite{milajerdi2019holmes,han2020unicorn,abdulellah2021atlas,yu2021alchemist,li2024nodlink} demonstrate that whole-system audit logs can recover complex multi-stage attacks. Subsequent work has improved scalability and graph analysis through systems such as Flash, Kairos, DistDet, PROGRAPHER, and ProvG-Searcher~\cite{10646725,10646673,feng2023distdet,fanyang2023prographer,altinisik2023provg}. Other approaches recognize tactics and techniques online or detect lateral movement through temporal graph analysis~\cite{lv2024trec,zhang2025tapas,qiao2025slot,aly2025ocrapt,king2023euler,khoury2024jbeil}.

Although these systems reconstruct attack behavior with high fidelity, they derive behavioral structure from execution telemetry rather than from structured CTI. ATT\&CK primarily serves as an annotation vocabulary, while the procedural information is recovered from observed execution. Our work instead investigates the opposite direction: how much procedural structure already exists before any execution takes place.

Adversary emulation frameworks assume executable procedures rather than descriptive intelligence. Systems such as Laccolith~\cite{orbinato2024laccolith}, PentestGPT~\cite{deng2024pentestgpt}, PentestAgent~\cite{shen2025pentestagent}, InCALMO~\cite{singer2025incalmo}, and Perry~\cite{singer2025perry} execute attacks through explicitly defined procedures that specify execution order, parameters, environmental assumptions, cleanup operations, and target-specific bindings. Likewise, planning frameworks such as AURORA~\cite{wang2024sands}, Effects Language~\cite{damodaran2025automated}, KNOWHOW~\cite{meng2025knowhow}, and MultiKG~\cite{wang2024multikg} rely on explicit procedural models that extend beyond descriptive CTI.

Related standardization efforts pursue similar goals. OASIS CACAO~\cite{cacao2spec} defines a standardized representation for executable security playbooks, while MITRE Attack Flow~\cite{attackflowctid} explicitly models ordered ATT\&CK procedures. These efforts recognize that executable cyber operations require information that is not directly represented in current ATT\&CK-in-STIX datasets.

Our work is complementary to these systems. Rather than proposing another execution language or planning framework, we ask a more fundamental question: how much of the required procedural information can already be recovered from public ATT\&CK-in-STIX before any analyst-assisted translation, planning model, or execution framework is introduced? To answer this question, we treat the ATT\&CK Enterprise bundle itself as the object of measurement and systematically characterize the procedural semantics that are present, missing, and therefore must be reconstructed before structured CTI becomes executable.

\section{Behavioral Metrics}
\label{sec:definitions}

Our objective is to quantify the extent to which reusable procedural information is encoded in public ATT\&CK-in-STIX before any analyst-assisted reconstruction. To accomplish this, we model campaigns and intrusion sets through the techniques they reference and measure the structural properties of the resulting behavioral representation space. These measurements characterize whether public CTI contains sufficient behavioral diversity and consistency to serve as the foundation for reusable adversary emulation.

The ATT\&CK Enterprise bundle represents adversary knowledge as a heterogeneous graph of typed STIX objects. Among the available object types, our analysis focuses on \texttt{attack-pattern}, \texttt{campaign}, \texttt{intrusion-set}, and the \texttt{relationship} objects that connect them. Additional entities, including \texttt{malware}, \texttt{tool}, \texttt{course-of-action}, and \texttt{x-mitre-tactic}, provide complementary behavioral context, while auxiliary objects such as \texttt{indicator}, \texttt{identity}, \texttt{report}, and \texttt{note} primarily contribute descriptive metadata. Table~\ref{tab:sources} summarizes the publicly available STIX sources incorporated into our consolidated dataset.

Although these objects describe adversary behavior, they do not explicitly encode executable procedures. ATT\&CK techniques and sub-techniques are represented as \texttt{attack-pattern} objects, with hierarchical relationships captured through the \texttt{sub-technique-of} relation and tactic assignments specified through the \texttt{kill\_chain\_phases} field. However, procedural information, including execution order, operational dependencies, environmental assumptions, parameter bindings, and control flow, is not represented in a machine-readable structure. Instead, these elements remain embedded in free-text descriptions and therefore cannot be directly recovered through structural analysis alone.

To evaluate the behavioral information that \emph{is} explicitly represented, we construct a binary campaign-technique matrix in which each campaign is represented by a vector indicating the presence or absence of every ATT\&CK technique. This representation enables quantitative comparison across campaigns while remaining independent of analyst interpretation or procedural reconstruction.

We evaluate four complementary structural properties of this matrix. Together, these measurements quantify the breadth of documented adversary behavior, the density of campaign descriptions, the degree of behavioral reuse between campaigns, and the existence of reusable procedural structure.

\begin{textbox}
\textbf{Behavioral Metrics.}

\textbf{Coverage:} Fraction of ATT\&CK techniques and sub-techniques observed at least once across all campaigns.

\textbf{Sparsity:} Proportion of zero entries in the campaign-technique matrix, measuring how completely campaigns describe adversary behavior.

\textbf{Overlap:} Pairwise similarity between campaigns based on their shared techniques, indicating the degree of behavioral reuse.

\textbf{Clustering:} Structural cohesion of campaign representations is measured through the silhouette coefficient, indicating whether campaigns naturally form reusable behavioral groups.
\end{textbox}
\begin{figure*}[!t]
    \centering
    \resizebox{\textwidth}{!}{
\usetikzlibrary{arrows.meta,calc,shapes.geometric}%
\providecolor{execblue}{HTML}{4477AA}%
\providecolor{ctired}{HTML}{EE6677}%
\providecolor{loopgreen}{HTML}{228833}%
\providecolor{midgray}{HTML}{666666}%
\providecolor{dashgray}{HTML}{BBBBBB}%
\providecolor{edgegray}{HTML}{333333}%
\providecolor{facegray}{HTML}{FFFFFF}%
\providecolor{stagefill}{HTML}{F5F5F5}%
\providecommand{\stix}{STIX}%
\begin{tikzpicture}[
  x=1cm,y=1cm,
  font=\sffamily\footnotesize,
  hex/.style={
    shape=regular polygon,
    regular polygon sides=6,
    shape border rotate=30,
    minimum size=2.04cm,
    text width=1.60cm,
    inner sep=1.8pt,
    align=center,
    line width=0.80pt,
    fill=edgegray!6,
    draw=edgegray!50,
    text=edgegray!96!black,
    font=\sffamily\bfseries\fontsize{7.20}{7.60}\selectfont
  },
  hexfirst/.style={hex, fill=execblue!10, draw=execblue!56, text=execblue!88!black},
  hexexec/.style={hex, fill=execblue!11, draw=execblue!62, text=execblue!92!black},
  arrow/.style={
    -{Latex[length=5.0pt,width=4.1pt]},
    line width=1.48pt,
    draw=edgegray!76
  },
  toplabel/.style={
    font=\sffamily\bfseries\fontsize{7.45}{7.9}\selectfont,
    text=edgegray!94!black,
    align=center,
    text width=3.08cm,
    inner sep=0pt
  },
  statelab/.style={
    font=\sffamily\bfseries\fontsize{7.30}{7.8}\selectfont,
    align=center,
    inner sep=0pt,
    text=loopgreen!82!black
  },
  analystlab/.style={
    font=\sffamily\bfseries\fontsize{7.30}{7.8}\selectfont,
    align=center,
    inner sep=0pt,
    text=ctired!80!black
  }
]

\def\yHex{0}
\def\xFirst{1.08}
\def\xStep{3.42}

\coordinate (c1) at (\xFirst,\yHex);
\coordinate (c2) at ({\xFirst+\xStep},\yHex);
\coordinate (c3) at ({\xFirst+2*\xStep},\yHex);
\coordinate (c4) at ({\xFirst+3*\xStep},\yHex);
\coordinate (c5) at ({\xFirst+4*\xStep},\yHex);

\node[hexfirst] (cti)  at (c1) {Structured\\CTI};
\node[hex, fill=edgegray!10, draw=edgegray!56] (s1) at (c2) {Stage 1:\\Structural\\Modeling};
\node[hex, fill=ctired!18, draw=ctired!74, text=ctired!95!black] (s2) at (c3) {Stage 2:\\Technique\\Translation};
\node[hex, fill=edgegray!10, draw=edgegray!56] (s3) at (c4) {Stage 3:\\Emulation\\Integration};
\node[hexexec] (exec) at (c5) {Executable\\Adversary\\Emulation};

\draw[arrow] ([xshift=2pt]cti.east) -- ([xshift=-2pt]s1.west);
\draw[arrow] ([xshift=2pt]s1.east)  -- ([xshift=-2pt]s2.west);
\draw[arrow] ([xshift=2pt]s2.east)  -- ([xshift=-2pt]s3.west);
\draw[arrow] ([xshift=2pt]s3.east)  -- ([xshift=-2pt]exec.west);

\node[toplabel] (ann1) at ($(cti.north)+(0,0.88cm)$)  {Behavioral entities\\in \stix{} bundles};
\node[toplabel] (ann2) at ($(s1.north)+(0,0.88cm)$)   {Normalize CTI into\\typed behavioral graph};
\node[toplabel] (ann3) at ($(s2.north)+(0,0.88cm)$)   {Explicit procedural\\specification};
\node[toplabel] (ann4) at ($(s3.north)+(0,0.88cm)$)   {Deploy and\\orchestrate plans};
\node[toplabel] (ann5) at ($(exec.north)+(0,0.88cm)$) {Reproducible\\multi-stage emulation};

\draw[line width=0.34pt, draw=edgegray!22]
  ($(ann1.south)!0.53!(cti.north)+(-0.36cm,0)$) --
  ($(ann5.south)!0.53!(exec.north)+(0.36cm,0)$);

\node[statelab]   (auto1) at ($(s1.south)+(0,-0.56cm)$) {Fully automated};
\node[analystlab] (auto2) at ($(s2.south)+(0,-0.56cm)$) {Analyst-supervised};
\node[statelab]   (auto3) at ($(s3.south)+(0,-0.56cm)$) {Scripted};

\node[
  font=\sffamily\bfseries\fontsize{7.10}{7.55}\selectfont,
  text=ctired!78!black,
  inner sep=0pt,
  align=center
] (gaptext) at ($(auto2.south)+(0,-0.15cm)$) {procedural gap};

\draw[line width=0.70pt, draw=ctired!46, line cap=round]
  ($(gaptext.west)+(-0.26cm,0)$) -- ($(gaptext.west)+(-0.07cm,0)$);
\draw[line width=0.70pt, draw=ctired!46, line cap=round]
  ($(gaptext.east)+(0.07cm,0)$) -- ($(gaptext.east)+(0.26cm,0)$);

\end{tikzpicture}%
}
    \caption{Overview of the proposed methodology for transforming descriptive CTI into executable multi-stage adversary behavior. The pipeline separates information directly recoverable from structured CTI from the procedural knowledge that must be supplied during translation.}
    \label{fig:methodology-overview}
\end{figure*}

\section{Methodology}
\label{sec:methodology}

Our objective is to determine whether structured CTI can be systematically transformed into executable adversary behavior while explicitly distinguishing the information recovered from CTI from the procedural knowledge supplied during translation. To this end, we propose a three-stage methodology that separates structural analysis, procedural reconstruction, and execution. The methodology is intentionally independent of any specific CTI representation or emulation framework. It requires only structured behavioral entities and explicit relationships among them. Throughout this paper, we instantiate the methodology using ATT\&CK-in-STIX as the CTI source and MITRE CALDERA as the execution platform.

Figure~\ref{fig:methodology-overview} summarizes the proposed workflow. Stage~1 automatically extracts and normalizes behavioral knowledge from structured CTI. Stage~2 reconstructs the procedural semantics that are absent from the original representation by translating behavioral descriptions into executable actions. Stage~3 packages these actions into an adversary emulation environment, enabling controlled execution and validation. Together, these stages establish a reproducible boundary between descriptive intelligence and executable procedure.

\subsection{Stage 1: Structural Modeling}
\label{sec:stage1}
The objective of the first stage is to construct a uniform behavioral representation from structured CTI. Independent of the original serialization, threat entities are normalized into a typed graph whose vertices represent behavioral objects, including techniques, campaigns, intrusion sets, malware, and tools, and whose edges represent semantic relationships such as \texttt{uses}, \texttt{attributed-to}, and \texttt{subtechnique-of}. This representation preserves the behavioral information explicitly encoded by the CTI source while remaining independent of any execution platform.

To support quantitative analysis, campaigns and intrusion sets are projected into a common behavioral representation space defined by ATT\&CK techniques. Each campaign is represented as a binary vector indicating the presence or absence of every technique. Since structured CTI rarely specifies temporal order, we derive a deterministic, tactic-aligned sequence by assigning each technique to its earliest ATT\&CK tactic, following the canonical progression from Reconnaissance to Impact. This ordering is not interpreted as an execution trace; instead, it provides a stable organizational scaffold for comparing campaigns and evaluating structural similarity.

The output of Stage~1 consists of a normalized behavioral graph, campaign-technique matrices, and tactic-aligned technique sequences that provide the structural foundation for the remaining stages.

\subsection{Stage 2: Procedural Translation}
\label{sec:stage2}
Stage~2 addresses the gap in procedural semantics identified throughout this paper. While Stage~1 captures what adversaries are reported to have done, it does not determine how those behaviors should be executed. Consequently, each technique must be translated into an executable action that is consistent with both the original CTI description and the target execution environment.

Translation proceeds over the tactic-aligned technique sequence produced during Stage~1. For each technique, we first recover the behavioral intent encoded in the CTI representation. Existing public implementations, such as Atomic Red Team, may be reused when available, while language models can assist in generating candidate commands when no implementation exists. Regardless of the source, every translated action is validated against the original CTI description and the declared System Under Test (SUT).

During this process, all information absent from the structured CTI, including parameter values, privilege assumptions, environmental bindings, execution conditions, and validation logic, is explicitly documented rather than inferred implicitly. The output of Stage~2 is therefore not only an executable action sequence but also a traceable record distinguishing procedural knowledge recovered from CTI from knowledge introduced during translation.

ShadowRay illustrates this process. ATT\&CK identifies technique T1190 (Exploit Public-Facing Application) and associates it with abuse of exposed Ray services. However, the structured representation does not specify the vulnerable endpoint, request parameters, or success criteria required for execution. These elements are introduced only during Stage~2 through explicit analyst validation and environment binding. Consequently, the executable command becomes a reproducible implementation of the behavioral intent rather than a direct encoding of the original CTI.

\subsection{Stage 3: Emulation Integration}
\label{sec:stage3}
The final stage converts translated actions into executable adversary workflows. Unlike the previous stages, which remain independent of any execution platform, Stage~3 depends on the capabilities provided by the selected emulation framework. Each translated action is represented as an executable ability together with its execution requirements, expected outcome, and cleanup behavior. Ordered collections of abilities form adversary profiles that approximate the behavioral progression described by the original CTI.

In this work, we instantiate Stage~3 using MITRE CALDERA. Translated actions are mapped to CALDERA abilities and assembled into adversary profiles executed through automated operations. Once the translated actions have been validated during Stage~2, the remaining execution pipeline, including artifact generation, profile construction, orchestration, and repeated experimentation, is fully automated.

The methodology therefore establishes a clear separation of responsibilities. Stage~1 automatically recovers behavioral structure from CTI. Stage~2 reconstructs the procedural semantics required for execution under analyst supervision. Stage~3 operationalizes the resulting workflow within an adversary emulation platform. This separation makes explicit which components of an executable campaign originate from structured CTI and which necessarily result from procedural reconstruction.

To demonstrate the methodology, we present two representative case studies derived from the public ATT\&CK-in-STIX \emph{ShadowRay} and \emph{Soft Cell} datasets. Their translated workflows are executed in a reproducible, Docker-based CALDERA environment to evaluate whether structured CTI can support coherent, multi-stage adversary emulation once the missing procedural semantics are supplied. Rather than reproducing historical intrusions exactly, these experiments quantify the practical boundary between descriptive CTI and executable adversary behavior.

\section{Behavioral Analysis}
\label{sec:analysis}

We begin by asking a fundamental question: \emph{how much procedural information is already encoded in publicly available ATT\&CK-in-STIX?} If structured CTI captures reusable adversary behavior, campaigns should collectively provide broad behavioral coverage, exhibit consistent structural organization, and expose recurring execution patterns. Conversely, if campaign descriptions primarily document isolated observations, the resulting behavioral representation space should remain sparse, fragmented, and weakly reusable.

To answer this question, we first characterize the behavioral content of the public ATT\&CK Enterprise bundle. We then compare campaign and intrusion-set representations to determine whether either provides a stable procedural abstraction. Finally, we evaluate whether the observed behavioral structure naturally forms reusable execution patterns suitable for adversary emulation.

The ATT\&CK Enterprise bundle represents the richest publicly available source of structured adversary behavior currently distributed in STIX format. Unlike most public CTI repositories, it simultaneously models campaigns, intrusion sets, techniques, malware, tools, and their semantic relationships within a single knowledge base. Consequently, it provides the strongest available baseline for evaluating the procedural sufficiency of structured CTI.

Although our consolidated corpus includes multiple public STIX repositories, quantitative measurements in this section are intentionally limited to the Enterprise bundle. The remaining sources primarily contribute ecosystem-level diversity and frequently contain incomplete campaign descriptions, heterogeneous schemas, or duplicated behavioral information. If a reusable procedural structure cannot be identified in Enterprise, it is unlikely to emerge consistently from sparser public datasets.

Our first measurement evaluates behavioral coverage. Campaign objects reference only \campaignCoveragePct\% of the available techniques, indicating that most documented adversary behavior never appears in any public campaign object. Intrusion sets cover a substantially larger portion of the ATT\&CK technique space because they aggregate observations across multiple operations. However, this broader coverage reflects behavioral accumulation rather than procedural completeness.

Coverage alone, therefore, does not imply executability. Campaigns identify behaviors observed during operations, while intrusion sets summarize long-term adversary capabilities. Neither representation specifies how those behaviors should be ordered, parameterized, or adapted to a concrete execution environment. Structured CTI therefore captures substantial behavioral knowledge but only a partial description of executable adversary procedures.

\subsection{Campaigns and Intrusion Sets Capture Different Behavioral Views}
\label{sec:campaign_intrusionsets}

Campaigns and intrusion sets serve different purposes within ATT\&CK. Campaigns describe individual operations performed during a bounded period of activity, whereas intrusion sets aggregate recurring behaviors observed across multiple campaigns. If either representation were sufficiently complete to enable adversary emulation, it should expose a stable behavioral organization that can be reused across operations.

To compare these representations, we construct binary behavioral vectors that indicate the presence or absence of each ATT\&CK technique and project them into a common behavioral representation space. Figure~\ref{fig:intrusionset_pca} visualizes this projection using principal component analysis.

The resulting distributions reveal clear differences between the two object types. Campaigns occupy highly dispersed regions of the behavioral representation space, reflecting the operational diversity and incompleteness of individual CTI reports. In contrast, intrusion sets form denser behavioral profiles because they accumulate techniques observed across multiple campaigns. This aggregation produces richer behavioral summaries but simultaneously obscures the procedural structure of individual operations.

The asymmetry between campaigns and intrusion sets also exposes an important limitation of public CTI. Campaign objects frequently omit operational phases, intermediate actions, or environmental context, while intrusion sets intentionally abstract away temporal distinctions in favor of long-term behavioral characterization. Consequently, neither representation directly captures the ordered sequence of actions required to emulate the adversary.

This observation is consistent with previous measurements of public CTI quality, which report heterogeneous reporting practices, incomplete behavioral descriptions, and inconsistent use of ATT\&CK annotations across providers~\cite{jin2024sharing,10117505,vermeer2022evolution}. Our results extend these findings by showing that the structural limitations persist even within the most comprehensive publicly available ATT\&CK dataset.

\begin{figure}[t]
    \centering
\usetikzlibrary{calc,decorations.pathreplacing}%
\providecolor{execblue}{HTML}{4477AA}%
\providecolor{ctired}{HTML}{EE6677}%
\providecolor{loopgreen}{HTML}{228833}%
\providecolor{edgegray}{HTML}{333333}%
\providecolor{midgray}{HTML}{666666}%
\providecolor{dashgray}{HTML}{BBBBBB}%
\providecolor{facegray}{HTML}{FFFFFF}%
\providecolor{stagefill}{HTML}{F5F5F5}%
\providecommand{\nTechniquesAnalysis}{691}%
\providecommand{\campaignTechniqueCount}{297}%
\providecommand{\campaignCoveragePct}{43.0}%
\providecommand{\intrusionSetTechniqueCount}{488}%
\providecommand{\intrusionSetCoveragePct}{70.6}%
\providecommand{\campaignOnlyCount}{29}%
\providecommand{\campaignOnlyPct}{4.2}%
\providecommand{\sharedTechniqueCount}{268}%
\providecommand{\sharedPct}{38.8}%
\providecommand{\intrusionSetOnlyCount}{220}%
\providecommand{\intrusionSetOnlyPct}{31.8}%
\providecommand{\unusedTechniqueCount}{174}%
\providecommand{\unusedPct}{25.2}%
\begin{tikzpicture}[font=\sffamily, every node/.style={outer sep=0pt}]

\def\W{8.46}
\def\H{1.10}
\def\N{691}

\pgfmathsetmacro{\wCampaignOnly}{\campaignOnlyCount/\N*\W}
\pgfmathsetmacro{\wShared}{\sharedTechniqueCount/\N*\W}
\pgfmathsetmacro{\wIntrOnly}{\intrusionSetOnlyCount/\N*\W}
\pgfmathsetmacro{\wUnused}{\unusedTechniqueCount/\N*\W}

\pgfmathsetmacro{\xA}{0}
\pgfmathsetmacro{\xB}{\wCampaignOnly}
\pgfmathsetmacro{\xC}{\wCampaignOnly+\wShared}
\pgfmathsetmacro{\xD}{\wCampaignOnly+\wShared+\wIntrOnly}
\pgfmathsetmacro{\xE}{\W}

\fill[ctired!78!white] (\xA,0) rectangle (\xB,\H);
\fill[ctired!42!execblue!55!white] (\xB,0) rectangle (\xC,\H);
\fill[execblue!80!white] (\xC,0) rectangle (\xD,\H);
\fill[dashgray!48] (\xD,0) rectangle (\xE,\H);
\draw[edgegray!92!black, line width=0.58pt] (\xA,0) rectangle (\xE,\H);
\foreach \x in {\xB,\xC,\xD}{
  \draw[white, line width=0.42pt] (\x,0) -- (\x,\H);
}

\node[
  font=\sffamily\bfseries\fontsize{7.9}{8.3}\selectfont,
  text=edgegray!96!black,
  anchor=south,
  inner sep=0pt
] at ({\W/2},\H+1.14) {Total techniques: \nTechniquesAnalysis};

\draw[ctired!72!black, line width=0.64pt]
  (\xA,\H+0.62) -- (\xC,\H+0.62);
\draw[ctired!72!black, line width=0.50pt] (\xA,\H+0.53) -- (\xA,\H+0.71);
\draw[ctired!72!black, line width=0.50pt] (\xC,\H+0.53) -- (\xC,\H+0.71);
\node[
  font=\sffamily\fontsize{6.80}{7.15}\selectfont\bfseries,
  text=ctired!86!black,
  anchor=south,
  align=center,
  inner sep=0pt
] at ({(\xA+\xC)/2},\H+0.78)
  {\campaignTechniqueCount\ campaign techniques (\campaignCoveragePct\%)};

\draw[execblue!72!black, line width=0.64pt]
  (\xB,\H+0.18) -- (\xD,\H+0.18);
\draw[execblue!72!black, line width=0.50pt] (\xB,\H+0.10) -- (\xB,\H+0.26);
\draw[execblue!72!black, line width=0.50pt] (\xD,\H+0.10) -- (\xD,\H+0.26);
\node[
  font=\sffamily\fontsize{7.25}{7.65}\selectfont\bfseries,
  text=execblue!86!black,
  anchor=south,
  align=center,
  inner sep=0pt,
  text width=4.90cm
] at ({(\xB+\xD)/2},\H+0.24)
  {\intrusionSetTechniqueCount\ intrusion-set techniques (\intrusionSetCoveragePct\%)};

\node[font=\sffamily\bfseries\fontsize{8.6}{8.9}\selectfont, text=edgegray!98!black]
  at ({(\xA+\xB)/2},0.49*\H) {\campaignOnlyCount};
\node[font=\sffamily\bfseries\fontsize{9.8}{10.2}\selectfont, text=edgegray!98!black]
  at ({(\xB+\xC)/2},0.49*\H) {\sharedTechniqueCount};
\node[font=\sffamily\bfseries\fontsize{9.8}{10.2}\selectfont, text=edgegray!98!black]
  at ({(\xC+\xD)/2},0.49*\H) {\intrusionSetOnlyCount};
\node[font=\sffamily\bfseries\fontsize{8.6}{8.9}\selectfont, text=edgegray!98!black]
  at ({(\xD+\xE)/2},0.49*\H) {\unusedTechniqueCount};

\node[
  font=\sffamily\fontsize{7.15}{7.55}\selectfont,
  text=edgegray!92!black,
  align=left,
  text width=1.30cm,
  anchor=north west,
  inner sep=0pt
] at (\xA,-0.16) {Campaign only\\(\campaignOnlyPct\%)};
\node[
  font=\sffamily\fontsize{7.15}{7.55}\selectfont,
  text=edgegray!92!black,
  align=center,
  text width=0.96cm,
  anchor=north
] at ({(\xB+\xC)/2},-0.16) {Shared\\(\sharedPct\%)};
\node[
  font=\sffamily\fontsize{7.15}{7.55}\selectfont,
  text=edgegray!92!black,
  align=center,
  text width=1.78cm,
  anchor=north
] at ({(\xC+\xD)/2},-0.16) {Intrusion-set only\\(\intrusionSetOnlyPct\%)};
\node[
  font=\sffamily\fontsize{7.15}{7.55}\selectfont,
  text=edgegray!92!black,
  align=right,
  text width=1.05cm,
  anchor=north east,
  inner sep=0pt
] at (\xE,-0.16) {Unused\\(\unusedPct\%)};

\end{tikzpicture}%
    \caption{Behavioral projection of campaigns and intrusion sets in the ATT\&CK Enterprise bundle. Campaigns occupy dispersed regions of the behavioral representation space, whereas intrusion sets aggregate behaviors across multiple operations, thereby forming denser profiles. Neither representation exposes the procedural ordering required for executable emulation of adversaries.}
    \label{fig:intrusionset_pca}
\end{figure}

These observations motivate the remaining analyses in this section. Coverage quantifies how much adversary behavior is documented, while the campaign-intrusion-set comparison demonstrates that a broader behavioral representation does not necessarily imply procedural completeness. We next investigate whether the naturally available behavioral information forms reusable execution patterns through clustering and sequence-based analyses.

\subsection{Behavioral Structure Does Not Form Reusable Campaign Families}
\label{sec:clustering}

The previous analysis shows that campaigns and intrusion sets capture different views of adversary behavior. We next ask whether the documented behaviors nevertheless organize into reusable structural families. If campaigns consistently follow similar behavioral patterns, their technique representations should naturally cluster into cohesive groups.

To evaluate this hypothesis, we apply unsupervised clustering to the campaign-technique matrix constructed in Section~\ref{sec:definitions}. Campaigns are represented by binary vectors over the ATT\&CK technique space, and cluster quality is evaluated using the silhouette coefficient. Higher silhouette values indicate well-separated behavioral groups, whereas values near zero imply weak structural organization.

Across all tested values of $k$, clustering quality remains consistently low. The best silhouette coefficient is between 0.03 and 0.05. These results indicate that campaigns do not partition into well-defined behavioral families. Instead, they exhibit gradual variation with substantial overlap between neighboring campaigns.

This outcome is consistent with the descriptive nature of public CTI. Individual campaigns typically document only the techniques observed during a particular investigation, and the level of detail varies considerably across reports. Consequently, the resulting behavioral vectors capture partial observations rather than complete procedural workflows.

To ensure that these observations were not artifacts of the clustering configuration, we repeated the analysis using multiple initialization strategies and parameter settings. The resulting silhouette values remained consistently low, indicating that the absence of coherent campaign families reflects the underlying behavioral representation rather than the clustering algorithm itself.

The lack of natural clustering has an important implication for adversary emulation. If campaigns do not form stable behavioral groups, reusable adversary profiles cannot be obtained simply by partitioning existing campaign objects. Additional procedural information must therefore be introduced during translation rather than inferred from campaign similarity alone.

\subsection{Campaigns Do Not Share a Common Procedural Backbone}
\label{sec:lcs}

Weak clustering alone does not preclude the existence of a reusable execution structure. Campaigns might still share a common sequence of techniques even if their overall behavioral profiles differ substantially. We therefore investigate whether public CTI encodes a stable procedural backbone that appears across multiple campaigns.

Because ATT\&CK does not explicitly encode execution order, we first construct deterministic tactic-aligned sequences as described in Section~\ref{sec:stage1}. These sequences provide a common organizational scaffold while avoiding assumptions about temporal execution. We then compare campaign pairs using the Longest Common Subsequence (LCS), which measures the longest ordered sequence of shared techniques.

The resulting LCS values are consistently small. Most campaign pairs share only short partial behavioral patterns, and no sequence appears frequently enough to constitute a reusable execution template. Even campaigns attributed to the same intrusion set often diverge substantially in their observed technique sequences.

To determine whether these short subsequences arise from genuine behavioral organization or from the overall popularity of common ATT\&CK techniques, we compare the observed results against randomized baselines. Randomized campaigns produce LCS distributions comparable to those observed in the original dataset, indicating that much of the apparent overlap can be explained by technique frequency rather than by shared procedural structure.

These observations reinforce the conclusions drawn from the clustering analysis. Public CTI captures recurring behavioral elements, but those elements do not combine into stable execution chains that can be directly reused for adversary emulation. Instead, campaigns appear as collections of partially overlapping behavioral patterns whose procedural relationships remain implicit.

The absence of a common behavioral backbone reflects the intended role of ATT\&CK. The framework records evidence that particular techniques have been observed during an intrusion, but it does not attempt to reconstruct the operational sequence connecting those observations. Consequently, structured CTI provides a description of adversary behavior without encoding the executable procedure necessary for replay.

Taken together, the clustering and LCS analyses demonstrate that the limitations of ATT\&CK-in-STIX extend beyond incomplete coverage. Even where behavioral information is available, it rarely forms reusable procedural structures suitable for direct execution. This finding motivates the translation methodology proposed in Section~\ref{sec:methodology}, in which missing procedural semantics are reconstructed explicitly rather than inferred from statistical regularities in the public dataset.

\section{Emulation Setup}
\label{sec:emulation}

The objective of our experimental environment is to evaluate whether CTI-derived workflows can execute consistently once the missing procedural semantics have been reconstructed. The testbed is therefore designed as a controlled, reproducible execution environment rather than a faithful reproduction of the original victim infrastructures. This distinction is important because our goal is to validate the procedural sufficiency of the translated workflows, not to recreate historical intrusions.

Following the three-stage methodology, Stage~1 automatically extracts campaigns, intrusion sets, and ATT\&CK techniques from the Enterprise STIX bundle and organizes them using the conservative tactic-aligned scaffold introduced in Section~\ref{sec:methodology}. Stage~2 translates each technique into an executable action through analyst validation and explicitly documents every parameter, environmental assumption, privilege requirement, and execution dependency introduced during reconstruction.

Stage~3 operationalizes these translated actions using MITRE CALDERA~5.3.0. In our implementation, CALDERA functions solely as an execution engine. It neither interprets STIX objects nor infers sequencing, parameter values, or procedural relationships. Instead, every translated technique is explicitly represented as a CALDERA ability that includes the validated command and optional cleanup logic. Abilities are subsequently assembled into adversary profiles according to the behavioral scaffold generated during Stage~1. Consequently, execution failures reflect inconsistencies in the reconstructed procedure or unmet environmental assumptions rather than inference performed by the emulation framework itself.

\begin{figure}[htpb]
    \centering
    \includegraphics[width=0.99\linewidth]{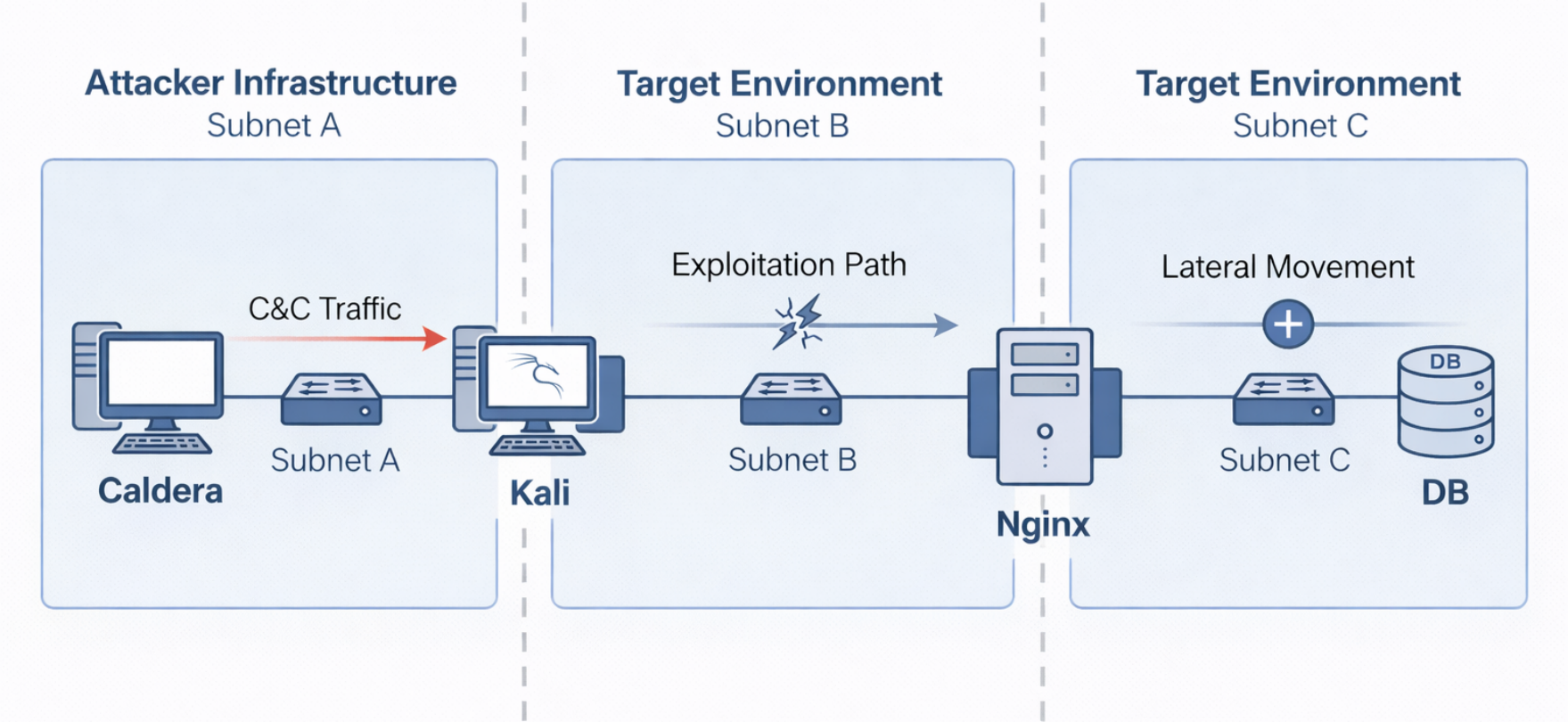}
    \caption{Experimental environment used to evaluate CTI-derived workflows. The left side contains the execution infrastructure, consisting of the \caldera{} server and a Kali execution agent connected through a command-and-control channel. The right side represents the shared laboratory System Under Test (SUT), where campaign-specific services and vulnerabilities are instantiated to execute workflows.}
    \label{fig:env}
\end{figure}

Experiments were conducted using the legacy Docker artifact distributed with CALDERA, which provides a curated collection of eight representative adversary workflows together with the supporting attacker and target services. As summarized in Table~\ref{tab:campaign_techniques}, this environment consists of a CALDERA server, a Kali Linux execution agent, and multiple target services interconnected through isolated Docker networks. Within this architecture, CALDERA orchestrates the execution workflow, while the Kali agent performs the translated adversary actions, consistent with the conceptual architecture illustrated in Figure~\ref{fig:env}.

The System Under Test (SUT) consists of four Docker containers: one Kali Linux~2025.4 container acting as the adversary host and three Debian~13 containers representing the target infrastructure. During image construction, each target container is configured with the vulnerabilities and software components required for the translated ATT\&CK techniques associated with the evaluated campaigns. This build-time configuration ensures that every experimental run begins from an identical execution state, thereby eliminating configuration drift between repetitions.

Within the SUT, one Debian container exposes the public-facing services required for initial access, while the remaining hosts provide the internal services necessary for post-compromise activity, privilege escalation, and lateral movement. Campaigns are represented as ordered collections of ATT\&CK-aligned CALDERA abilities executed by the Kali agent against the designated targets. Execution begins only after the environment has completed its initialization sequence, ensuring that all environmental assumptions established during Stage~2 are satisfied before the first translated action is issued.

The experimental environment is intentionally minimal. Rather than approximating the complexity of operational enterprise networks, it provides a stable execution environment for evaluating whether CTI-derived workflows remain internally consistent once the missing procedural semantics have been supplied explicitly. This controlled design isolates the translation methodology from unrelated infrastructure variability and enables reproducible evaluation across repeated executions.

To facilitate reproducibility, the complete implementation, including the translated workflows, Docker environment, automation scripts, and execution artifacts, will be released as open source after the conclusion of the double-blind review process.\footnote{Repository URL withheld during double-blind review and released upon publication.}

\begin{table}[ht]
\centering
\caption{Number of ATT\&CK techniques per campaign in the case study}
\label{tab:campaign_techniques}
\begin{tabular}{l r}
\hline
\textbf{Campaign} & \textbf{\# Techniques} \\[1pt]
\hline
APT41 DUST & 23 \\
C0010 & 9 \\
C0026 & 6 \\
CostaRicto & 10 \\
Operation MidnightEclipse & 17 \\
Outer Space & 8 \\
Salesforce Data Exfiltration & 18 \\
ShadowRay & 10 \\
\hline
\textbf{Total} & \textbf{101} \\
\hline
\end{tabular}
\end{table}

Operationally, each enactment experiment requires four components: (i) a reproducible execution environment, (ii) the campaign-specific services and vulnerabilities required by the translated techniques, (iii) the reconstructed workflow produced by the translation methodology, and (iv) execution traces that record the resulting behavior. The legacy Docker artifact satisfies these requirements by providing a shared laboratory environment onto which campaign-specific bootstrap scripts are applied. Rather than provisioning an independent System Under Test (SUT) for each campaign, all translated workflows execute over the same controlled substrate. Consequently, campaigns that historically involved multiple hosts, such as \emph{Soft Cell}, are evaluated here as procedural enactments within a common experimental environment. The objective is not to reproduce the original operational infrastructure, but to determine whether the reconstructed workflow executes consistently once the required procedural semantics are supplied.

Figure~\ref{fig:env} summarizes the experimental architecture. The environment deliberately minimizes infrastructure complexity so that execution outcomes reflect the translated workflow rather than incidental properties of the underlying platform. By controlling the execution environment across all experiments, differences in execution behavior can be attributed to the reconstructed procedures instead of environmental variability.

\begin{figure*}[htpb]
    \centering
    \resizebox{0.99\textwidth}{!}{
\usetikzlibrary{arrows.meta,calc,positioning,shapes.geometric}%
\providecolor{execblue}{HTML}{4477AA}%
\providecolor{ctired}{HTML}{EE6677}%
\providecolor{loopgreen}{HTML}{228833}%
\providecolor{midgray}{HTML}{666666}%
\providecolor{dashgray}{HTML}{BBBBBB}%
\providecolor{edgegray}{HTML}{333333}%
\providecolor{facegray}{HTML}{FFFFFF}%
\providecolor{stagefill}{HTML}{F5F5F5}%
\providecommand{\gapkey}[1]{{\sffamily\fontsize{7.0}{7.4}\selectfont\textcolor{midgray!92!black}{#1}}}%
\begin{tikzpicture}[
  font=\sffamily,
  panel/.style={
    rounded corners=1.9pt,
    line width=0.64pt,
    minimum width=\PanelW,
    minimum height=\PanelH,
    fill=white
  },
  stixpanel/.style={panel, draw=execblue!58},
  execpanel/.style={panel, draw=ctired!58},
  bridgehex/.style={
    shape=regular polygon,
    regular polygon sides=6,
    shape border rotate=30,
    minimum size=2.04cm,
    text width=1.60cm,
    inner sep=1.8pt,
    align=center,
    draw=ctired!74,
    fill=ctired!18,
    line width=0.80pt,
    text=ctired!95!black,
    font=\sffamily\bfseries\fontsize{7.20}{7.60}\selectfont
  },
  bridgearrow/.style={
    -{Latex[length=5pt,width=3.8pt]},
    line width=0.76pt,
    draw=edgegray!54
  },
]

\def\TotalW{15.70cm}
\def\PanelW{5.90cm}
\def\PanelH{4.95cm}
\def\PanelTextW{5.08cm}
\def\PanelTextH{3.88cm}
\def\PanelPadX{0.18cm}
\def\PanelPadY{0.17cm}

\node[stixpanel, anchor=north west] (stix) at (0,0) {};
\node[
  anchor=north west,
  inner sep=0pt,
  align=left,
  font=\sffamily\fontsize{8.0}{8.6}\selectfont
] at ($(stix.north west)+(\PanelPadX,-\PanelPadY)$) {%
  \parbox[t][\PanelTextH][t]{\PanelTextW}{%
    \textcolor{execblue!86!black}{\textbf{Structured CTI provides}}\\[3.2pt]
    {\gapkey{Technique}}\\[-1pt]
    \textbf{T1190} -- Exploit Public-Facing Application\\[4pt]
    {\gapkey{Tactic phase}}\\[-1pt]
    Initial Access\\[4pt]
    {\gapkey{Platform}}\\[-1pt]
    Linux, macOS, Windows, \ldots\\[4pt]
    {\gapkey{Description (free text)}}\\[-1pt]
    \textit{``Adversaries may exploit vulnerabilities in internet-facing applications\ldots''}\\[4pt]
    {\gapkey{Campaign context}}\\[-1pt]
    ShadowRay -- abuse of exposed Ray dashboard services%
  }%
};

\node[execpanel, anchor=north east] (exec) at (\TotalW,0) {};
\node[
  anchor=north west,
  inner sep=0pt,
  align=left,
  font=\sffamily\fontsize{8.0}{8.6}\selectfont
] at ($(exec.north west)+(\PanelPadX,-\PanelPadY)$) {%
  \parbox[t][\PanelTextH][t]{\PanelTextW}{%
    \textcolor{ctired!86!black}{\textbf{Executable step requires}}\\[3.2pt]
    {\gapkey{Target address}}\\[-1pt]
    \texttt{172.21.0.20}\\[4pt]
    {\gapkey{Port \& endpoint}}\\[-1pt]
    \texttt{5055/exec}\\[4pt]
    {\gapkey{HTTP method \& payload}}\\[-1pt]
    \texttt{POST}, multipart form data\\[4pt]
    {\gapkey{Validation logic}}\\[-1pt]
    Check response contains expected output\\[4pt]
    {\gapkey{Concrete command}}\\[-1pt]
    \texttt{curl -X POST -F 'cmd=whoami'}\\
    \texttt{http://172.21.0.20:5055/exec}%
  }%
};

\node[bridgehex] (bridge) at ($(stix.center)!0.5!(exec.center)$) {%
  Stage 2\\[-0.10pt]
  {\fontsize{5.9}{6.2}\selectfont\textcolor{midgray!92!black}{analyst-supervised}}%
};

\draw[bridgearrow] (stix.east |- bridge.center) -- (bridge.west);
\draw[bridgearrow] (bridge.east) -- (exec.west |- bridge.center);

\node[
  font=\sffamily\fontsize{7.0}{7.4}\selectfont,
  text=edgegray!96!black,
  inner sep=0pt,
  text width=14.10cm,
  align=center,
  below=0.28cm of $(stix.south)!0.5!(exec.south)$
] (gaplabel) {%
  \textbf{Procedural gap:} structured CTI omits parameters, environment bindings,
  privilege assumptions, protocol details, and success criteria.
  The analyst must supply them before execution.%
};

\end{tikzpicture}%
}
    \caption{Illustration of the gap in procedural semantics using technique T1190 from the ShadowRay case study. The left panel shows the behavioral information provided by ATT\&CK-in-STIX, whereas the right panel summarizes the additional procedural information introduced during Stage~2. Concrete parameters, environment bindings, privilege assumptions, and validation logic are reconstructed explicitly because they are not represented in the original structured CTI.}
    \label{fig:gap-example}
\end{figure*}

Figure~\ref{fig:gap-example} illustrates the role of the translation methodology using a representative technique from the ShadowRay campaign. While ATT\&CK identifies the behavioral objective associated with technique T1190, execution additionally requires concrete parameters, environment bindings, privilege assumptions, and validation criteria. These elements are not inferred by the execution framework; they are explicitly introduced during Stage~2 while preserving the behavioral intent encoded in the original CTI.

Each experiment follows the same evaluation protocol. Structured CTI is first transformed into a tactic-aligned behavioral scaffold (Stage~1), after which individual techniques are translated into executable actions (Stage~2). The translated workflow is then instantiated within the Docker-backed CALDERA environment (Stage~3), executed after the SUT reaches a stable configuration, and monitored through execution logs and link-level traces. When repeated executions are required, the environment is rebuilt to restore an identical initial state, ensuring that workflow execution behavior remains directly comparable across runs.

Execution is evaluated using three observable metrics: the number of successfully completed links, whether execution reaches the explicit workflow termination marker, and whether any non-success link states remain after execution has stabilized. Because transient scheduler states may occur during execution, measurements are collected only after the operation has reached a quiescent state. Within the legacy artifact, workflow completion corresponds to the terminal \texttt{T1529} ability, while residual non-success links indicate incomplete execution chains.

Across the eight curated adversary workflows distributed with the legacy artifact, execution produces 109 successful links, with a median of 11 successful links per workflow (range: 7-24). All eight workflows reach their explicit termination marker, and none retain residual non-success links after execution stabilizes. These observations demonstrate that the curated procedures execute consistently within the shared laboratory environment.

Importantly, these results should not be interpreted as evidence that ATT\&CK-in-STIX directly encodes executable adversary procedures. The execution artifact automates environment preparation, workflow orchestration, and evidence collection, but it does not reconstruct procedural semantics from CTI. Every executable workflow evaluated in this study depends on analyst-supplied parameter bindings, environmental assumptions, and execution constraints introduced during Stage~2.

Taken together, the experimental results reinforce the central finding of this paper. Public ATT\&CK-in-STIX provides sufficient behavioral information to guide adversary reconstruction, but successful execution requires explicit procedural enrichment beyond the information contained in the original structured representation. Even within a controlled execution environment, prerequisites, parameterization, environment bindings, and conditional behavior remain external to the current ATT\&CK Enterprise STIX model.

\section{Case Studies}
\label{sec:casestudies}

The behavioral analyses presented in the previous sections establish that ATT\&CK-in-STIX captures adversary behavior but does not explicitly encode executable procedures. We now evaluate how far the proposed translation methodology can bridge this gap. Our objective is not to faithfully reproduce historical intrusions, but to determine whether structured CTI provides sufficient behavioral information to construct coherent multi-stage adversary workflows once the missing procedural semantics are explicitly supplied.

We begin by characterizing the translation envelope of the Enterprise corpus. Among the \nIntrusionSets{} intrusion sets contained in the ATT\&CK Enterprise bundle, \nIntrusionSetsWithTechniques{} (\intrusionSetWithTechniquesPct\%) are associated with at least one ATT\&CK technique, whereas \nIntrusionSetsEmpty{} (\intrusionSetEmptyPct\%) contain no technique relationships. Across the populated intrusion sets, we observe \intrusionSetTechRefs{} technique references, corresponding to an average of \intrusionSetTechAvg{} techniques per intrusion set. These populated intrusion sets define the behavioral representation space available for translation.

Applying Stage~2 to all \nTechniquesAnalysis{} active ATT\&CK techniques further bounds the extent to which structured CTI can be operationalized. Of these, \nPlatformAgnosticTechniques{} techniques represent platform-independent behaviors that cannot be expressed directly as operating-system commands. The remaining \nTranslatableTechniques{} techniques are successfully translated into single-step CALDERA abilities. This result establishes the representational coverage of the translation methodology rather than its execution fidelity, which is evaluated through the case studies below.

To examine procedural reconstruction in practice, we consider two complementary case studies. ShadowRay represents a campaign that can be executed directly within the legacy Docker artifact, allowing us to evaluate execution on an existing adversary workflow. Soft Cell represents a larger intrusion reconstructed from the GALLIUM intrusion set and illustrates the additional procedural assumptions required when campaigns span multiple hosts and operational phases.

\subsection{Case Study I: ShadowRay}

ShadowRay provides a representative example of the gap in procedural semantics for a workflow already supported by the legacy execution environment. The underlying ATT\&CK representation identifies the techniques associated with the campaign but does not specify the executable procedure connecting them. Following the methodology described in Section~\ref{sec:methodology}, each technique is translated into an executable CALDERA ability while preserving its behavioral intent.

Most discovery and enumeration techniques require only generic operating-system commands. In contrast, techniques associated with Initial Access, Persistence, and Privilege Escalation require explicit parameter binding, environment-specific service information, privilege assumptions, and validation logic that are absent from the structured CTI representation. These procedural elements are introduced explicitly during Stage~2 rather than inferred automatically by the execution framework.

Execution within the legacy Docker environment demonstrates that the translated workflow progresses through multiple successful execution links and reaches its explicit workflow termination marker. Successful execution therefore validates the reconstructed procedure rather than demonstrating that the original ATT\&CK representation was itself executable.

\subsection{Case Study II: Soft Cell}

Soft Cell provides a complementary evaluation involving a multi-host intrusion reconstructed from the GALLIUM intrusion set. Compared with ShadowRay, the campaign spans a larger portion of the ATT\&CK matrix and therefore exposes additional procedural dependencies.

Although ATT\&CK identifies the behavioral objectives associated with the intrusion, reconstructing an executable workflow requires assumptions regarding host allocation, privilege transitions, routing, environmental configuration, and execution ordering. These assumptions accumulate rapidly as the workflow extends across multiple systems, illustrating that procedural complexity grows substantially faster than behavioral complexity.

Because CALDERA performs no automatic inference or workflow repair, every missing parameter, environmental dependency, or ordering constraint manifests directly as an incomplete execution plan. Soft Cell therefore highlights the practical boundary between descriptive CTI and executable adversary procedures.

\subsection{Generalization Across Legacy Adversaries}

The two focal case studies are supplemented by an execution audit of all eight curated adversary workflows distributed with the legacy Docker artifact. Together with the corpus-scale translation analysis, these experiments provide evidence at four complementary levels: (i) structural characterization of the ATT\&CK Enterprise corpus, (ii) technique-to-ability translation coverage, (iii) detailed procedural reconstruction through representative case studies, and (iv) execution behavior across a separate collection of curated adversaries.

Across all evaluated workflows, the same conclusion emerges consistently. Structured CTI provides the behavioral foundation necessary to reconstruct adversary activity, but executable workflows become possible only after explicit procedural semantics, including parameter bindings, environmental assumptions, execution ordering, and validation logic, are supplied during translation.

The case studies consistently reinforce the findings of the structural analysis. Public ATT\&CK-in-STIX captures the behavioral intent of adversary campaigns but does not encode the procedural semantics required for direct execution. Execution order, parameter bindings, privilege assumptions, environmental context, and control-flow decisions remain external to the current representation and must be reconstructed during translation.

Despite these limitations, the experiments demonstrate that structured CTI provides a sufficiently rich behavioral foundation for reproducible adversary emulation once the missing procedural semantics have been explicitly supplied. Across both focal case studies and the broader audit of the eight curated adversaries, translated workflows execute coherently within the controlled laboratory environment while remaining traceable to their originating CTI entities. These results therefore validate the proposed methodology and establish a clear boundary between the behavioral knowledge encoded by current CTI standards and the procedural knowledge that still needs to be introduced before automated adversary execution becomes possible.

\section{Discussion and Implications}
\label{sec:discussion}

The results presented throughout this paper consistently point to the same conclusion: public ATT\&CK-in-STIX captures the behavioral semantics of adversary campaigns but does not encode the procedural semantics required for direct execution. This distinction explains both the strengths and the limitations of current structured CTI. Behavioral entities, techniques, and relationships provide a rich description of adversary activity, yet executable workflows require additional information that lies outside the current STIX representation.

Our measurements show that the limitation of ATT\&CK-in-STIX is not primarily one of behavioral coverage. Although campaign objects describe only part of the ATT\&CK technique space, even complete behavioral coverage would not eliminate the gap in procedural semantics identified in this work. Execution order, parameter bindings, privilege assumptions, environmental context, branching logic, and success conditions are currently absent from the representation.

This observation also explains the apparent contrast between behavioral distinctiveness and executability. Campaigns may be distinguishable within the ATT\&CK corpus through characteristic combinations of techniques, yet those same campaigns remain procedurally incomplete. Likewise, our findings complement recent work on threat-group attribution~\cite{saha2026kitten}: identifying distinctive behavioral profiles is fundamentally different from reconstructing executable adversary procedures.

An important contribution of this work is the explicit identification of the automation boundary. Structural extraction from ATT\&CK-in-STIX can be fully automated because the required behavioral entities and relationships are explicitly represented in the data model. In contrast, procedural reconstruction requires information that is either absent or encoded only implicitly within free-text descriptions.

Our audit of structured ATT\&CK fields illustrates this distinction. Fields such as \texttt{kill\_chain\_phases} and \texttt{x\_mitre\_platforms} directly support Stage~1 of the methodology by providing tactic assignments and platform constraints. However, fields that would support executable procedures, including concrete commands, parameter values, privilege requirements, environmental bindings, execution guards, and success conditions, are either unavailable or unpopulated in the current Enterprise bundle. Consequently, Stage~2 cannot be automated through rule-based processing alone and necessarily requires external procedural knowledge.

This automation boundary should be viewed as a property of the current CTI representation rather than of the translation methodology. Improvements in future CTI standards could progressively reduce the amount of analyst intervention that is required for procedural reconstruction.

The case studies demonstrate that structured CTI nevertheless provides a valuable behavioral foundation for adversary emulation. Once the missing procedural semantics are reconstructed explicitly, translated workflows execute consistently while remaining traceable to their originating ATT\&CK entities.

This separation between behavioral and procedural knowledge provides practical benefits. Analysts can distinguish information derived directly from CTI from assumptions introduced during workflow construction, making the resulting emulations easier to validate, reproduce, and compare across environments. Rather than treating executable workflows as direct representations of historical campaigns, the methodology makes explicit which components originate from structured intelligence and which arise from procedural reconstruction.

The experiments also reinforce the importance of environmental context. Real intrusions depend on network topology, available services, privilege relationships, defensive controls, and operational constraints. Since these characteristics are not represented within current ATT\&CK-in-STIX bundles, realistic adversary emulation necessarily requires explicit System Under Test (SUT) modeling in addition to behavioral translation.

The gap in procedural semantics identified in this study suggests several opportunities for future research. One direction is the evolution of CTI standards themselves. Existing efforts such as CACAO playbooks~\cite{cacao2spec} and MITRE Attack Flow~\cite{attackflowctid} already introduce procedural concepts that complement the descriptive nature of ATT\&CK. Extending structured CTI with optional representations for execution order, environmental assumptions, parameter bindings, and control flow would substantially reduce the procedural burden currently imposed on downstream emulation frameworks.

A second direction concerns automated procedural reconstruction. Large Language Models (LLMs) offer a promising means of generating executable commands, environmental configurations, and workflow variants from descriptive CTI. In our experiments, LLM-generated procedures successfully executed 38 of the 101 evaluated techniques (37.6\%) without analyst intervention. Most unsuccessful cases resulted from minor configuration or parameterization errors rather than incorrect behavioral reasoning. After manual correction, execution success increased substantially, indicating that current limitations arise primarily from low-level grounding rather than from an inability to infer behavioral intent.

Finally, standardized benchmark datasets pairing structured CTI, narrative reports, executable procedures, and environment descriptions would enable systematic evaluation of future translation systems. Such resources would support reproducible comparisons among rule-based approaches, LLM-assisted translators, and emerging machine-actionable representations of CTI.

Overall, this work should be viewed as a boundary-mapping study. Rather than proposing a new CTI standard, it identifies precisely which behavioral information is already available in public ATT\&CK-in-STIX and which procedural semantics still need to be introduced before adversary behavior becomes directly executable in an operational environment. Making this distinction explicit provides a principled foundation for future work on machine-actionable cyber threat intelligence.

\section{Threats to Validity}
\label{sec:validity}

Like any empirical study of public CTI, our work is subject to several limitations. We discuss the principal threats to construct, internal, and external validity.

Our measurements characterize the procedural information explicitly represented in ATT\&CK-in-STIX rather than the complete behavior of real intrusions. Campaign-technique vectors, coverage, sparsity, overlap, clustering, and sequence analyses quantify the structured behavioral information available in public CTI, but they do not capture temporal execution, causal dependencies, branching behavior, failed attempts, or operational decision-making. Likewise, the tactic-aligned ordering introduced in Stage~1 serves only as a deterministic organizational scaffold and should not be interpreted as a reconstruction of the historical execution sequence.

This limitation is inherent to the current STIX representation and is consistent with the objective of this work, namely measuring the procedural sufficiency of structured CTI rather than reconstructing historical attack timelines.

Internal validity primarily depends on the correctness and consistency of the underlying ATT\&CK data. Public CTI is manually curated, evolves continuously, and occasionally contains incomplete relationships or inconsistent annotations across releases. During our analysis, we observed minor discrepancies between the downloadable Enterprise STIX bundle and the ATT\&CK website, particularly in campaign-to-group relationships.

To reduce these risks, all STIX objects were parsed deterministically, normalized through a common processing pipeline, and validated by consistency checks over object counts, identifiers, and relationships. Quantitative analyses were performed exclusively on the Enterprise bundle used throughout the study, ensuring that all reported measurements were computed from a single reproducible dataset.

Execution experiments introduce an additional source of internal validity. The legacy Docker artifact relies on a shared laboratory substrate that requires limited runtime preparation, including service initialization, CALDERA configuration, and architecture-specific adjustments before execution begins. Consequently, the reported execution results should be interpreted as evidence of procedural enactment within this controlled environment rather than as historical reproductions of the original campaigns.

The generality of our conclusions is bounded by both the CTI source and the evaluation environment. Quantitative measurements focus exclusively on the ATT\&CK Enterprise STIX bundle because, among the public CTI collections we analyzed, it preserves the richest campaign-level behavioral structure. Our conclusions, therefore, characterize the current public Enterprise representation and should not be interpreted as automatically applying to proprietary intelligence sources or other ATT\&CK domains, such as Mobile or ICS.

Similarly, the emulation environment was designed to evaluate procedural coherence rather than faithfully reproduce operational infrastructures. Because current STIX representations omit environmental assumptions, parameter bindings, privilege requirements, and execution constraints, these elements must be supplied during translation. Different analysts may therefore produce alternative yet executable realizations that are behaviorally consistent with the same structured CTI.

Finally, the two case studies and the evaluated adversaries represent a curated subset of workflows distributed with the legacy artifact. The reported execution characteristics should therefore be interpreted as evidence supporting the identified procedural boundary rather than as statistical estimates representative of all ATT\&CK campaigns or future CTI datasets.

\section{Conclusion}
\label{sec:conclusion}

This paper investigated whether the public ATT\&CK-in-STIX contains sufficient procedural information to support executable emulation of adversaries. Through a combination of structural measurement and end-to-end case studies, we showed that current structured CTI captures adversary behavior but does not encode the procedural semantics required for direct execution. Ordering, parameter bindings, environmental assumptions, and execution logic remain outside the current representation and must be reconstructed before workflows become executable. To make this boundary explicit, we proposed a three-stage methodology that separates automated structural modeling from analyst-supervised procedural translation and execution. The evaluation demonstrates that structured CTI provides a reproducible behavioral foundation for adversary emulation once the missing procedural semantics have been explicitly supplied. Overall, our findings establish a clear distinction between descriptive and executable cyber threat intelligence. We hope this work provides a foundation for future machine-actionable representations of CTI that combine standardized behavioral knowledge with explicit procedural semantics, enabling more reproducible and automated threat-informed defense.

\bibliographystyle{IEEEtran}
\bibliography{references}


\appendix

\section{Ethical Considerations}
The study focuses on assessing and enhancing adversary emulation as part of defensive cybersecurity research. Even though the methodology in question converts threat intelligence into executable emulation procedures, all experiments were conducted in controlled, isolated laboratory environments designed to avoid interaction with or impact on external systems.

The techniques and methods examined in this study have been obtained from publicly available MITRE ATT\&CK and STIX information and are used solely to reproduce adversary behavior for measurement, validation, and the engineering of detection systems. The experimental infrastructure was intentionally kept separate from the production networks, and the emulated activities were limited to the systems that the researchers controlled.

The generated commands and system configurations were used to determine whether threat-intelligence representations contain sufficient information to support the emulation of executable adversaries; they were not intended to enable unauthorized access, persistence, data exfiltration, or disruption of actual systems. Specifically, the experiments were designed to assess the semantic and environmental gaps between publicly available threat intelligence and executable procedures rather than to develop any new offensive capabilities.

The fact that the study uses publicly documented adversary techniques and emulation frameworks means the dual-use character of the resulting artifacts is recognized. To minimize the risks involved, publicly released artifacts are accompanied by documentation explaining the intended research context and the experimental assumptions. The experiments did not use any real-world credentials, private information, or data obtained from compromised systems.

The research therefore seeks to contribute to reproducible defensive security research by measuring the limitations of existing threat-intelligence representations and offering a controlled method for translating such information into adversary-emulation scenarios that can be experimentally verified.

\section{Open Science}

To enable reproducibility and allow for further research, we make the artifacts developed and used in this study available through an anonymized repository. These artifacts consist of the source code for the STICK framework, the scripts used for processing and analyzing MITRE ATT\&CK STIX data, and the components employed to generate and evaluate Caldera adversary-emulation artifacts.

The repository also includes the datasets and intermediate results needed in order to carry out the quantitative analyses described in this paper, such as the extraction and mapping of ATT\&CK techniques, campaigns, and intrusion sets; it provides configuration files together with documentation that describes the experimental environment and the way in which the experiments were carried out to aid in the reproduction of the experiments.

The artifact has been arranged so that researchers are able to carry out the main stages of the methodology on their own: (i) processing the ATT\&CK STIX representation, (ii) constructing the corresponding emulation artifacts, (iii) configuring the System Under Test (SUT), and (iv) carrying out and assessing the resulting adversary-emulation scenarios.

The repository includes all items that can be redistributed; third-party datasets, such as the MITRE ATT\&CK data, are cited in accordance with their original sources and are not redistributed if their licenses or terms do not allow redistribution.

The anonymized artifact repository is available at:

\url{https://anonymous.4open.science/r/sticks-5BC2/README.md}

Once the paper has been accepted, the repository will be made public.

\section*{LLM usage considerations}\label{sec:ai-tools}

The authors used Grammarly and ChatGPT for grammar checking, limited editorial revision, and minor formatting assistance for figures and illustrations. All scientific claims, analyses, experiments, visual representations, and conclusions were produced and verified by the authors.

\section{Automation-Relevant Field Population}
\label{app:field-population}

Table~\ref{tab:field_population} summarizes the structured-field audit over the \nAppendixAttackPatterns{} active Enterprise attack patterns used in this study. The pattern is consistent with the automation boundary discussed in Section~\ref{sec:discussion}: Stage~1 fields are populated, while the candidate Stage~2 fields are absent or empty in the current export.

\begin{table}[htbp]
\centering
\caption{Population of automation-relevant fields across the active Enterprise attack patterns.}
\label{tab:field_population}
\small
\begin{tabular}{lcc}
\toprule
\textbf{Field} & \textbf{Present} & \textbf{Non-empty} \\
\midrule
\AppendixFieldPopulationRows
\bottomrule
\end{tabular}
\end{table}

\section{Supplementary Set-Based Reuse Probe}
\label{app:itemset-probe}

To complement the tactic-ordered LCS analysis with an order-free probe, we compute for each itemset size $k \in \{1,\dots,5\}$ the maximum support of any $k$-technique itemset across the 51 Enterprise campaigns. The rapid decay in Table~\ref{tab:itemset_probe} reinforces the same conclusion as the clustering and LCS results: campaigns share small recurring fragments, but not a dominant reusable backbone.

\begin{table}[htbp]
\centering
\caption{Maximum support of any unordered technique itemset across the 51 Enterprise campaigns.}
\label{tab:itemset_probe}
\small
\begin{tabular}{ccc}
\toprule
\textbf{Itemset size} & \textbf{Max support} & \textbf{Campaign share} \\
\midrule
\AppendixItemsetRows
\bottomrule
\end{tabular}
\end{table}
\end{document}